\documentclass[sn-mathphys,iicol]{sn-jnl}

\jyear{2021}%

\theoremstyle{thmstyleone}%
\theoremstyle{thmstyletwo}%

\theoremstyle{thmstylethree}%

\begin{document}

\title{The total neutron production from the alpha induced reaction on natural Zirconium }


\author[1]{\fnm{Vafiya Thaslim T T} \sur{}}

\author*[1]{\fnm{M M Musthafa} \sur{}}\email{mmm@uoc.ac.in}

\author[1]{\fnm{Midhun C V} \sur{}}
\author[2]{S. Ghugre}
\author[1]{Gokul Das H}
\author[1]{Swapna B}
\author[1]{Najmunnisa T}
\author[3]{F S Shana}
\author[4]{Rijin N T}
\author[5]{S Dasgupta}
\author[5]{J Datta}

\affil*[1]{\orgdiv{Department of Physics}, \orgname{University of Calicut}, \city{Malappuram}, \postcode{673635}, \state{Kerala}, \country{India}}
\affil[2]{\orgdiv{UGC-DAE-Consortium for Scientific Research}, \city{Kolkata}, \postcode{700098}, \state{West Bengal}, \country{India}}
\affil[3]{\orgname{Govt. Arts and Science College}, \orgaddress{\street{Meenchanda}, \city{Kozhikode}, \postcode{673018}, \state{Kerala}, \country{India}}}

\affil[4]{\orgdiv{Department of Physics}, \orgname{Jain University}, \city{Banglore}, \postcode{560069}, \state{Karnataka}, \country{India}}
\affil[5]{\orgdiv{Analytical Chemistry Division}, \orgname{Bhabha Atomic Research Centre, Variable Energy Cyclotron Centre}, \city{Kolkata}, \postcode{700064}, \state{West Bengal}, \country{India}}


\abstract{A significant amount of alpha particles, upto $\sim$ 35 MeV are produced in the reactor environment. Alpha induced reaction on natural Zirconium, a reactor component, upto 40 MeV has been measured using stacked foil activation technique. The total neutron production cross section from all possible channels for $\alpha$ energies upto 35 MeV is also estimated using TALYS 1.96. The isomeric cross section ratio for the production of the radionuclide $^{95}Nb$ has been measured and reported for the first time.}




\maketitle

\section{Introduction}\label{sec1}

\begin{figure}[h]
\centering
\includegraphics[scale=0.25]{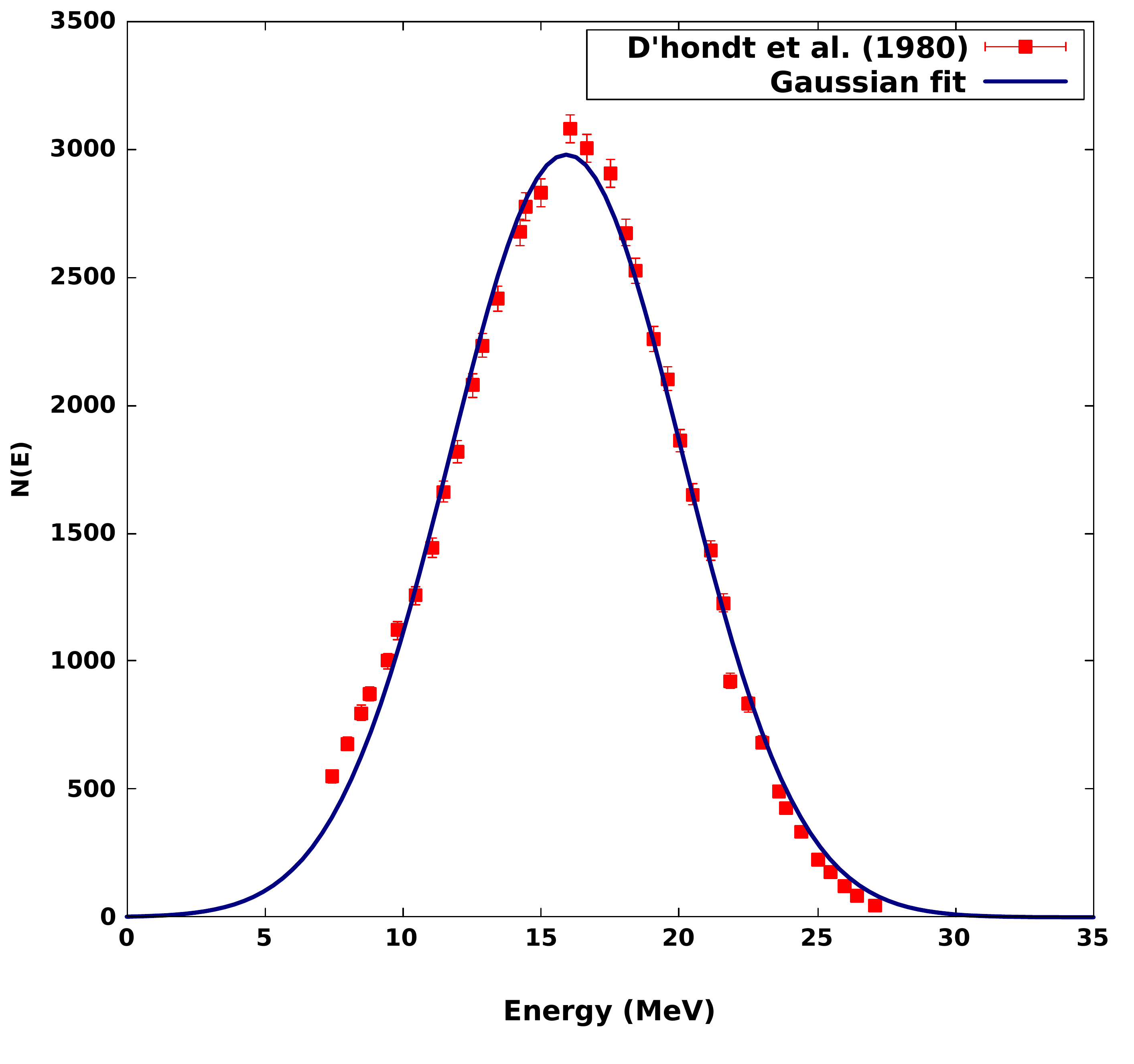}  
\caption{Energy distribution of alpha particles emitted in the $^{235}U(n,f)$ reaction.}
\label{distrbtn}
\end{figure}

Nuclear reaction studies in the vicinity of the reactor domain have the greatest scope even now. Possibilities of various nuclear reactions in the reactor environment and their effect should be accounted properly for the structural integrity and safe operation of critical systems. The spontaneous decay and ternary fission of heavy radioactive nuclei like U, Th, Pu, etc, used as fuel in the reactor, leading to the production of high energy alpha particles up to 35 MeV. Absolute yield of production of the alpha particles in the neutron-induced ternary fission of uranium nucleus is much more compared to the production of protons and tritons \cite{DHONDT1980461}. The energy distribution of alpha particles emitted in a typical neutron induced fission reaction on Uranium fluoride layers, with a thickness of 199 ${ \mu g}/{{cm}^2}$ and an isotopic enrichment of  99.524\% are shown in Fig. \ref{distrbtn} \cite{WAGEMANS2004291}. 
\par{ These alpha particles will induce various nuclear reactions with the materials used as the reactor components. 
Zirconium is one of the major components used as cladding or outer covering of the fuel rods, due to their low thermal neutron capture cross sections and high corrosion resistance. There is a chance for the interaction of the alpha particles with these Zirconium materials in the reactor domain within the energy range. Moreover, recent studies show the presence of alpha particles in the interested energy domain on the surface of the lunar farside from the Chang’E-4 mission \cite{Spacealpha}, this may lead to the interaction with Zirconium materials used as a structural material in spacecraft \cite{WEI2021150420}. The available experimental data for alpha induced reaction on natural Zirconium over the energies of interest are found to be limited in the EXFOR \cite{Otuka_2020} library and the literature. }

	Considering the above fact, we have studied the production cross-sections for  {$ ^{96}Zr(\alpha,x)^{99}Mo$}, {$ ^{nat}Zr(\alpha,x)^{93m}Mo$}, {$ ^{nat}Zr(\alpha,x)^{96}Nb $}, { $ ^{nat}Zr(\alpha,x)^{95m,g}Nb $}, {$ ^{nat}Zr(\alpha,x)^{92m}Nb $}, {$ ^{nat}Zr(\alpha,x)^{89g}Zr $} and {$ ^{nat}Zr(\alpha,x)^{90}Nb$} reaction channels. The residual production cross sections are determined and the production cross sections of the neutron, proton, deuteron, triton, and alpha particles are estimated for energy upto 35 MeV. To the best of our knowledge the production cross section for $^{nat}Zr(\alpha,x)^{90}Nb$ reaction in the interested energy domain is being reported for the first time. Also, the residual nucleus $^{89g}Zr$ and $^{90}Nb$ have high medical importance, particularly in the diagnosis of tumors and therapy.
\par{From the measured cross sections of { $^{nat}Zr(\alpha,x)^{95m,g}Nb $} reactions, the isomeric cross-section ratio(ICR) for {$^{95g,m}Nb$} residues is determined. To the best of our knowledge, the experimental isomeric cross-section ratio for the radionuclide {$^{95g,m}Nb$} is reported for the first time. Measurements on isomeric cross section ratio will give insight into nuclear structural studies, the population of the spin state of the nuclei \cite{Satheesh_2011,satheesh_2012} and progress of nuclear reactions. Theoretical analysis of the data has been carried out using TALYS 1.96 \cite{Talys} code.}
\section{Experimental Details and Analysis}\label{sec2}

\indent{The experiment was carried out at Variable Energy Cyclotron Center (VECC), Kolkata, India, using the K-130 Cyclotron facility. Stacked foil activation technique has been employed for the measurement of cross-sections. Two stacks containing the samples of natural Zirconium of thickness 6.49 $mg/{{cm}^2}$ were irradiated with alpha particles of 30 MeV and 40 MeV beams separately, with an average beam current of 100 nA. The four samples were arranged in two stacks to minimize the energy spread over the energy range of interest. The mean energy corresponding to each sample was $20.4$, $29.6$, $32.3$, and $39.7$ MeVs, and was calculated using SRIM \cite{SRIM}. The mean energy on each sample has been calculated on the basis of energy falling on the sample and the energy loss within the thickness of the sample. The activities induced in the samples were counted using 100cc  HPGe detector(CANBERRA) coupled to MCA. The detector was calibrated for energy and efficiency using standard point source of {$^{152}Eu$} of known activity. 

{The collected data were analyzed using the data analysis framework CANDLE \cite{candle} and the absolute photo peak efficiency, which depends upon the geometry is calculated by equation~(\ref{eqn1})
 \begin{center}
 \begin{equation}
 G(E_\gamma)=\frac{C(E_\gamma)}{{t} {A_0} {I_\gamma}}\label{eqn1}
 \end{equation}
 \end{center}
where $C(E_\gamma)$ is the count under the photo peak, t is the counting time, $A_0$ is the known activity of the {$^{152}Eu$} sample source and $I_\gamma$ is the branching ratio of the interested gamma ray. The production cross-sections of the obtained channels are determined by the known activation equation~(\ref{eqn2}) \cite{MUSTHAFA2005419}.
\begin{table}[h!]
\caption{The characteristics of the reaction products.}\label{Table1}
\begin{tabular}{llllr}

\hline
 \multicolumn{1}{p{0.7cm}}{\centering Residual Nucleus}&  \multicolumn{1}{p{0.7cm}}{\centering Half life} &  \multicolumn{1}{p{0.5cm}}{\centering Contributing\\ reactions} & \multicolumn{1}{p{0.5cm}}{\centering Q-value \\(MeV) }& \multicolumn{1}{p{0.5cm}}{\centering $E_{th}$\\ (MeV)}\\
\hline
$^{99}Mo$ & $65.97h $& $^{96}Zr(\alpha,n)$&$-5.12$&5.29\\
\hline
$^{93m}Mo$ & $6.85 h$& $^{90}Zr(\alpha,n)$&$-7.61$&7.95\\
           &         &$^{91}Zr(\alpha,2n)$&$-14.81$&15.46\\
           &         &$^{92}Zr(\alpha,3n)$&$-23.44$&24.46\\
\hline

$^{96}Nb$ & $23.35 h$& $^{94}Zr(\alpha,d)$&$-12.38$&12.9\\
		  &		 & $^{94}Zr(\alpha,pn)$&$-14.60 $&15.11\\
		  &          &$^{96}Zr(\alpha,tn)$&$-20.43$&21.28\\
		  &  			 &$^{96}Zr(\alpha,d2n)$&$-26.69 $&27.6\\
		  &			 &$^{96}Zr(\alpha,p3n)$&$-28.91 $&29.9\\		
		 			
\hline		  
$^{95g}Nb$ & $34.99 d $& $^{92}Zr(\alpha,p)$&$-6.54$&6.82\\
           &            &$^{94}Zr(\alpha,t)$&$-13.01$&13.56\\
           &		    &$^{94}Zr(\alpha,nd)$&$-19.27$&19.94\\
           &			&$^{94}Zr(\alpha,2np)$&$-21.49$&22.24\\
           &            &$^{96}Zr(\alpha,t2n)$&$-27.32$&28.46\\
          	&		    &$^{96}Zr(\alpha,3nd)$&$-33.58$&34.72\\
          	&			&$^{96}Zr(\alpha,4np)$&$-35.8$&37.02\\
           
\hline
$^{95m}Nb$ & $3.6 d $   & $^{92}Zr(\alpha,p)$&$-6.54$&6.82\\
           &            &$^{94}Zr(\alpha,t)$&$-13.01$&13.56\\
           &			&$^{94}Zr(\alpha,nd)$&$-19.27$&19.94\\
           & 			&$^{94}Zr(\alpha,2np)$&$-21.49$&22.24\\
           &            &$^{96}Zr(\alpha,2nt)$&$-27.32$&28.46\\
           &			&$^{96}Zr(\alpha,3nd)$&$-33.58$&34.72\\
           &			&$^{96}Zr(\alpha,4np)$&$-35.8$&37.02\\
            
         \hline
$^{92m}Nb$ & $10.15 d $& $^{90}Zr(\alpha,d)$&$-13.03$&13.61\\
 		   & 			& $^{90}Zr(\alpha,np)$&$-13.53$&15.82\\
           &            &$^{91}Zr(\alpha,t)$&$-13.97$&14.58\\
       	   &	 		&$^{91}Zr(\alpha,nd)$&$-20.22$&20.96\\
       	   &			&$^{91}Zr(\alpha,2np)$&$-22.45$&23.27\\
           &            &$^{92}Zr(\alpha,tn)$&$-22.6$&23.58\\
           &			&$^{92}Zr(\alpha,2nd)$&$-8.86$&29.9\\
           &			&$^{92}Zr(\alpha,3np)$&$-31.08$&32.2\\
          \hline
$^{90g}Nb$ & $14.6 h$	& $^{90}Zr(\alpha,tn)$&$-26.71$&27.9\\
			&			&$^{90}Zr(\alpha,2nd)$&$-32.96$&34.18\\
			&			& $^{90}Zr(\alpha,3np)$&$-35.19$&36.49\\
			&         &$^{91}Zr(\alpha,t2n)$&$-33.9$&$35.14$\\
         \hline       
$^{89g}Zr$ & $78.41 h $& $^{90}Zr(\alpha,\alpha n)$&$-11.97$&12.5\\

			&			& $^{90}Zr(\alpha,dt)$&$-29.56$&30.74\\
			&			& $^{90}Zr(\alpha,npt)$&$-31.78$&32.95\\
			&			& $^{90}Zr(\alpha,2n^{3}He)$&$-32.54$&33.75\\
			&			& $^{90}Zr(\alpha,n2d)$&$-35.81$&37.14\\

           &            &$^{91}Zr(\alpha,\alpha 2n)$&$-19.16$&20.01\\
           &				 &${91}Zr(\alpha,2t)$&$-30.49$&31.60\\
           &            &$^{92}Zr(\alpha,\alpha 3n)$&$-27.8$&29.01\\ 
           \hline           
             
\end{tabular}
\end{table}

\begin{table}[h!]
\caption{The measured cross sections of $\alpha$ induced reactions on $^{nat}Zr$.} \label{Table2}
\begin{center}
\begin{tabular}{llr}
\hline

 \multicolumn{1}{p{0.5cm}}{\centering Reaction} & \multicolumn{1}{p{0.5cm}}{\centering Energy\\(MeV)}  & \multicolumn{1}{p{1.5cm}}{\centering Cross section \\(mb)}\\
\hline
$ ^{96}Zr(\alpha,n)^{99}Mo$	&$20.4 \pm 0.5$&$84.25\pm 0.20$  \\
							&$29.6 \pm 0.4$&$24.64\pm 0.05$\\
							&$32.3 \pm 0.4$&$24.46\pm 0.05$ \\
							&$39.7 \pm 0.3$&\\
							\hline
$ ^{nat}Zr(\alpha,x)^{93m}Mo$ &$20.4 \pm 0.5$&$20.97\pm 0.38$\\
							&$29.6 \pm 0.4$&$47.31\pm 0.82$\\
							&$32.3 \pm 0.4$&$73.02\pm 1.43$\\
							&$39.7 \pm 0.3$& $106.67\pm 2.02$\\
							\hline 
 $ ^{nat}Zr(\alpha,x)^{96}Nb $	&$20.4 \pm 0.5$&$0.09\pm 0.01$\\
							&$29.6 \pm 0.4$&$5.52\pm 0.13 $\\
							&$32.3 \pm 0.4$&$8.26\pm 0.22$\\
							&$39.7 \pm 0.3$&$9.02\pm 0.33$\\
							\hline
 
  $ ^{nat}Zr(\alpha,x)^{95g}Nb $	&$20.4 \pm 0.5$&$ 1.43 \pm 0.05 $\\
							&$29.6 \pm 0.4$&$ 4.17 \pm 0.09$\\
							&$32.3 \pm 0.4$&$ 7.01 \pm  0.16$\\
							&$39.7 \pm 0.3$&$ 22.8 \pm 0.39$\\
							\hline
   $ ^{nat}Zr(\alpha,x)^{95m}Nb $	&$20.4 \pm 0.5$&$ 0.16\pm 0.03$\\
							&$29.6 \pm 0.4$&$ 0.61\pm 0.02 $\\
							&$32.3 \pm 0.4$&$ 0.79\pm 0.04$\\
							&$39.7 \pm 0.3$&$ 1.64\pm 0.09$\\
							\hline
   $ ^{nat}Zr(\alpha,x)^{92m}Nb $	&$20.4 \pm 0.5$&$ 0.66 \pm 0.05 $\\
							&$29.6 \pm 0.4$&$ 18.29 \pm 0.13$\\
							&$32.3 \pm 0.4$&$ 26.14 \pm 0.19$\\
							&$39.7 \pm 0.3$&$ 17.4 \pm 0.22$\\
							\hline
   $ ^{nat}Zr(\alpha,x)^{90}Nb$	&$20.4 \pm 0.5$&\\
							&$29.6 \pm 0.4$&  \\
							&$32.3 \pm 0.4$& $ 0.03\pm 0.01$\\
							&$39.7 \pm 0.3$&$ 1.35\pm 0.08 $\\
							\hline
   $ ^{nat}Zr(\alpha,x)^{89g}Zr $&$20.4 \pm 0.5$&$0.09  \pm 0.09  $\\
							&$29.6 \pm 0.4$&$  11.19\pm 0.08  $ \\
							&$32.3 \pm 0.4$&$22.28  \pm 0.11  $\\
							&$39.7 \pm 0.3$&$49.18  \pm 0.27  $\\
\hline
 \end{tabular}
\end{center}
\end{table} }
\begin{center}
\begin{equation}
\sigma(E_\gamma)=\frac{C(E_\gamma) \lambda e^{\lambda t_2}}{N_0 \Phi G(E_\gamma) I_\gamma (1-e^{-\lambda t_1})(1-e^{-\lambda t_3}) } \label{eqn2}
 \end{equation}\\
\end{center}
where $ \lambda$ is the decay constant of the residual nucleus, $ N_0$ is the number of target particles per unit area, $\Phi$ is the beam flux, $I_\gamma$ branching ratio of the gamma ray, $G(E_\gamma)$ is the geometry dependent efficiency of the detector for the given gamma energy $E_\gamma$, and $t_1$, $t_2$ and $t_3 $ are the irradiation time, the cooling time and the counting time respectively. Relevant data, for reaction channels contributing to the production of obtained residues over the measured range of energies are tabulated in Table \ref{Table1}. The measured cross sections of the residual nuclei formed through the $\alpha$ induced reactions on $^{nat}Zr$ are listed in Table \ref{Table2}. The uncertainty in the measurements \cite{PhysRevC.47.2055} is calculated by including the gamma counting error, error in half life, and gamma ray intensity, the last two were taken from updated IAEA data library \cite{nndc}.
 
\section{Theoretical Analysis}\label{sec3}
 The statistical model calculations for {$^{nat}Zr(\alpha,x)$ has been performed using nuclear reaction code Talys 1.96 \cite{Talys}. The calculations were done by considering the compound nuclear decay by Hauser Feshbach formalism. The production cross sections in TALYS code were calculated by taking into account all possible reaction channels based on their threshold energies. To get the theoretical cross sections for alpha on natural Zirconium target, we have incorporated the isotopic abundance of the individual stable isotopes of natural target. The isotopic abundances are listed in Table \ref{Table3}. The measured cross section at any energy is taken as the sum of the production from each contributing channel.
\begin{table}[h!]
\caption{The abundances of the isotopes present in  $^{nat}Zr$.}\label{Table3}
\centering

\begin{tabular}{lr}
\hline
Isotopes& \hspace{1mm} Abundance (\%) \\
\hline
$^{90}Zr$ &  $ 51.45 $\\
$^{91}Zr$ &  $ 11.22$\\
$^{92}Zr$ &  $ 17.15 $\\
$^{94}Zr$ &  $ 17.38$\\
$^{96}Zr$ &  $ 2.80$\\          
\hline
\end{tabular}

\end{table}

	In order to optimize the prediction of theoretical models, calculations have been performed using various macroscopic level density models, namely Constant temperature plus Fermi gas model(CTM) \cite{gcm}, Back shifted Fermi gas model(BFM) \cite{KONING200813}, Generalized superfluid model(GSM) \cite{{PhysRevC.47.1504}} and microscopic level density models, namely Skyrme Hartree Fock Bogoluybov level densities \cite{PhysRevC.5.626} and Gogny Hartree Fock Bogoluybov level densities \cite{PhysRevC.86.064317}, available in TALYS code. Further, different alpha optical model potentials were also used to get more accurate results, Global alpha potential by Avrigeanu et al. \cite{PhysRevC.90.044612} gives better reproduction of experimental data. A detailed study of the various reactions is done in the following sections.}
\subsection{{$^{nat}Zr(\alpha,n)^{99}Mo$} reaction}
\begin{figure}[h!]
\centering
\includegraphics[scale=0.25]{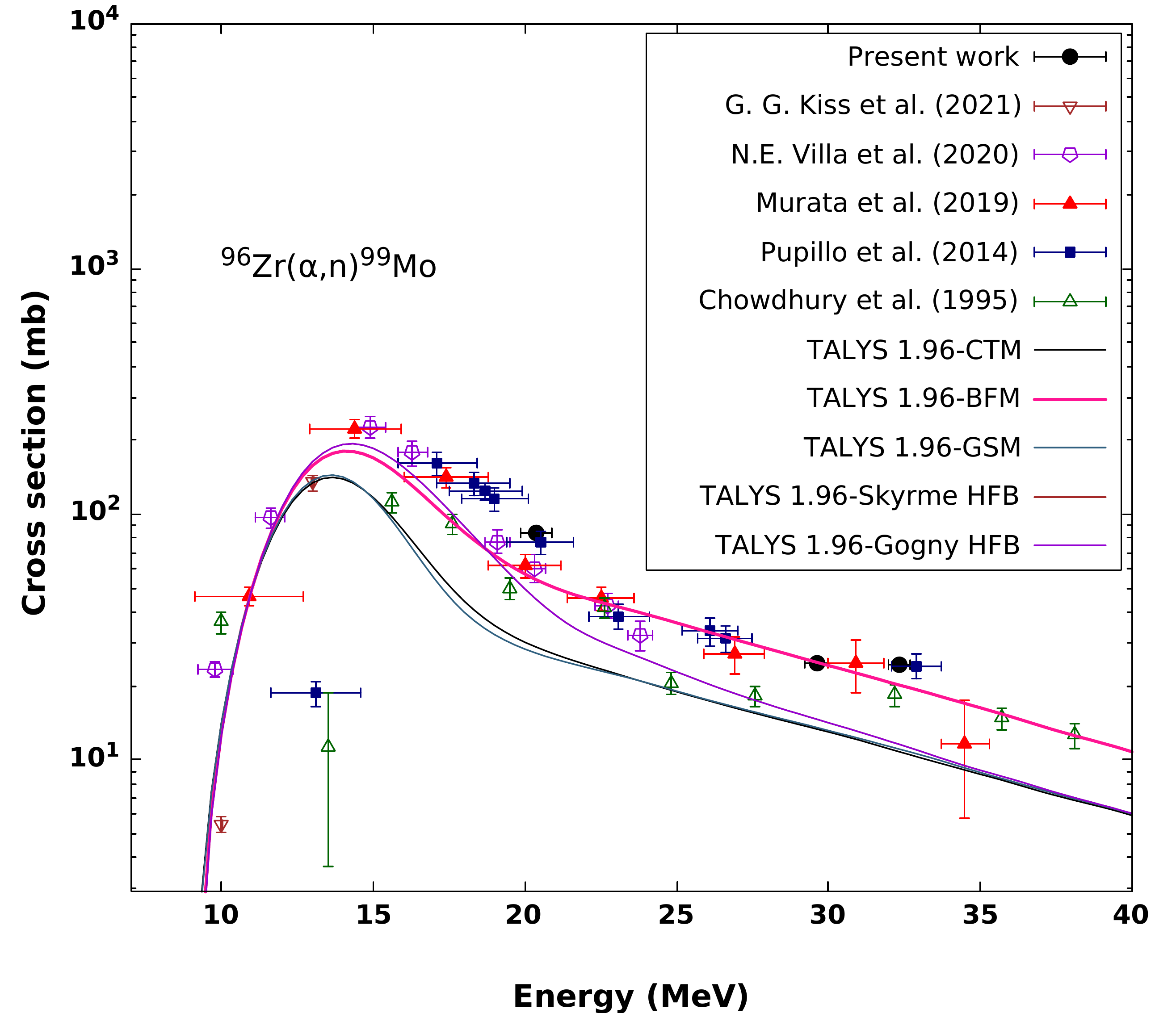} 
\caption{Experimentally measured excitation functions of the reaction  {$^{96}Zr(\alpha,n)^{99}Mo$}  with TALYS calculations.}\label{99Mo}
\end{figure}
{ The $^{99}Mo$ nucleus has a half life of 65.97 hours and is produced through the channel {$^{96}Zr(\alpha,n)^{99}Mo$}. The production cross section of  {$^{96}Zr(\alpha,n)^{99}Mo$} reactions for alpha beam of energies upto 40 MeV are measured using the activity of the gamma of energy 739.5 keV and is shown in the Fig. \ref{99Mo} along with available literature data and theoretical calculations. TALYS calculations with Back shifted Fermi gas model of level density show good agreement with present measurements. Measurements due to Chowdhury et al. \cite{chowdhury}, Pupillo et al. \cite{pupillo}, Murata et al. \cite{MUrata}, and N. E. Villa et al. \cite{villa} are also in good agreement with the present measurement. $^{99}Mo$ is an important radionuclide widely used as precursor for $^{99m}Tc$, used for medical imaging. Also low energy measurement of $^{96}Zr(\alpha,n)^{99}Mo$ reaction cross section is relevant in nucleosynthesis studies \cite{kiss_2021}.
\subsection{{$^{nat}Zr(\alpha,x)^{93m}Mo$} reaction}
\begin{figure}[h!]
\begin{center}
\includegraphics[scale=0.24]{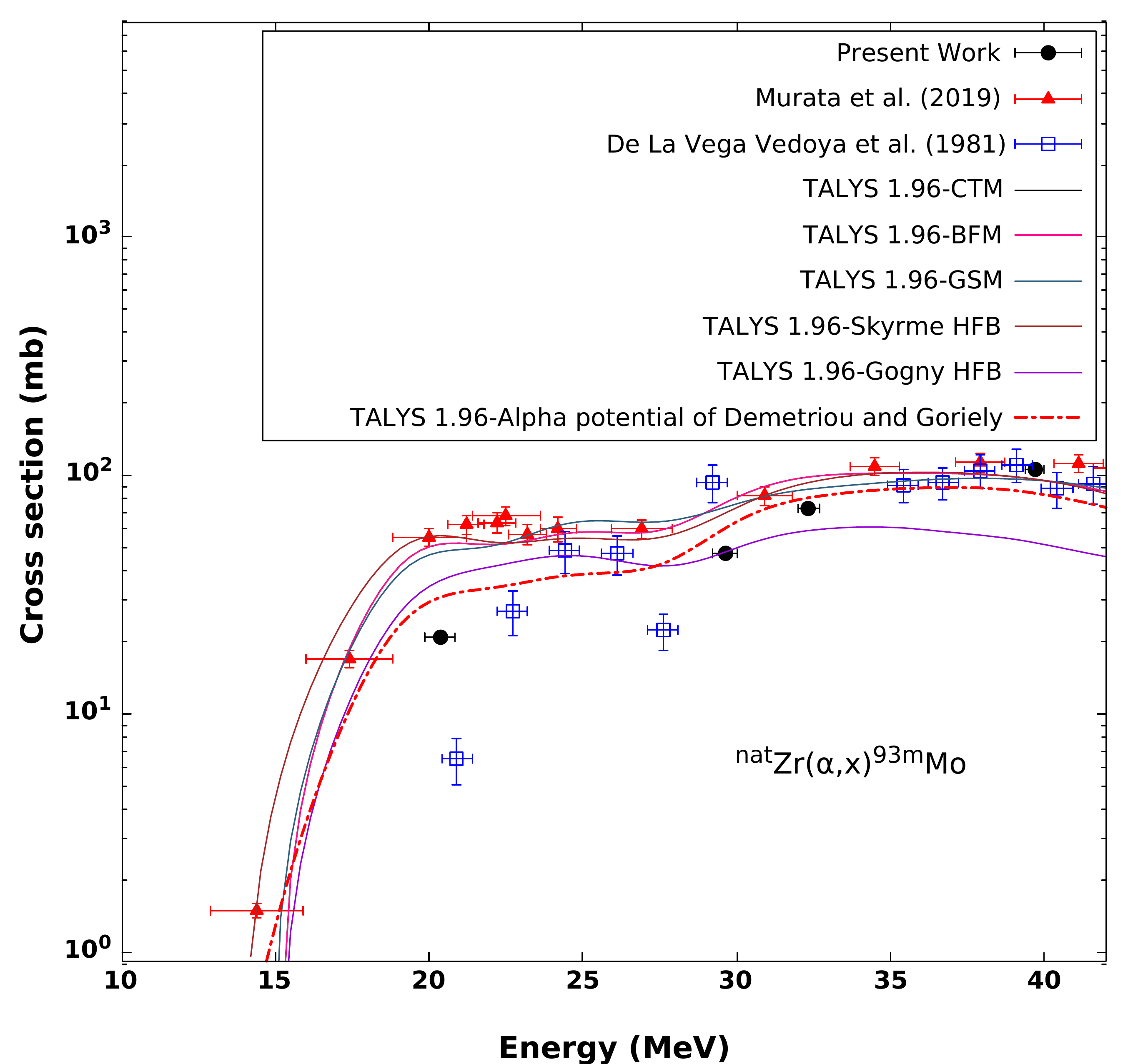}
\caption{Experimentally measured excitation functions of {$^{nat}Zr(\alpha,x)^{93m}Mo$} reaction with TALYS calculations.}\label{93mMo}
\end{center}
\end{figure}
{The radionuclide $^{93}Mo$ has an isomeric state $^{93m}Mo$ with a shorter half life of 6.85 hours, and ground state $^{93g}Mo$ with a longer half life of $4\times 10^3$ years. Production cross sections for {{$^{nat}Zr(\alpha,x)^{93g}Mo$} reaction can not be measured since the half life of the residual nucleus is very long. The major contributing channels are {$^{90}Zr(\alpha,n)^{93m}Mo$}, {$^{91}Zr(\alpha,2n)^{93m}Mo$}, and {$^{92}Zr(\alpha,3n)^{93m}Mo$}. Measured cross sections for the reaction {$^{nat}Zr(\alpha,x)^{93m}Mo$ are shown in Fig. \ref{93mMo}, the photo peak energy 263.049 keV, 684.693 keV, and 1477.138 keV are considered for the determination of production cross sections. Measurements by  De La Vega Vedoya et al. \cite{Vega_vedoya}, Murata et al., as well TALYS calculations are consistent with present measurements. However, at lower energies, the measurements of Murata et al. shows a larger discrepancy. The Gogny HFB model of TALYS calculations gives close agreement with present measurements at lower energies. Skyrme Hartree Fock Bogoluybov level density model and the other three macroscopic level density models shows similar trends at higher energies. Moreover, the dispersive optical model of Demetriou and Goriely gives good agreement with present measurements. In which the imaginary part of the potential is represented by Woods-Saxon potential where both surface and volume components are included by incorporating energy dependent terms, and a dispersion relation is used to connect both the real and imaginary parts of the potential \cite{Demetriou}.  
 \subsection{{$^{nat}Zr(\alpha,x)^{96}Nb$} reaction}
  \begin{figure}[h]
\centering
\includegraphics[scale=0.22]{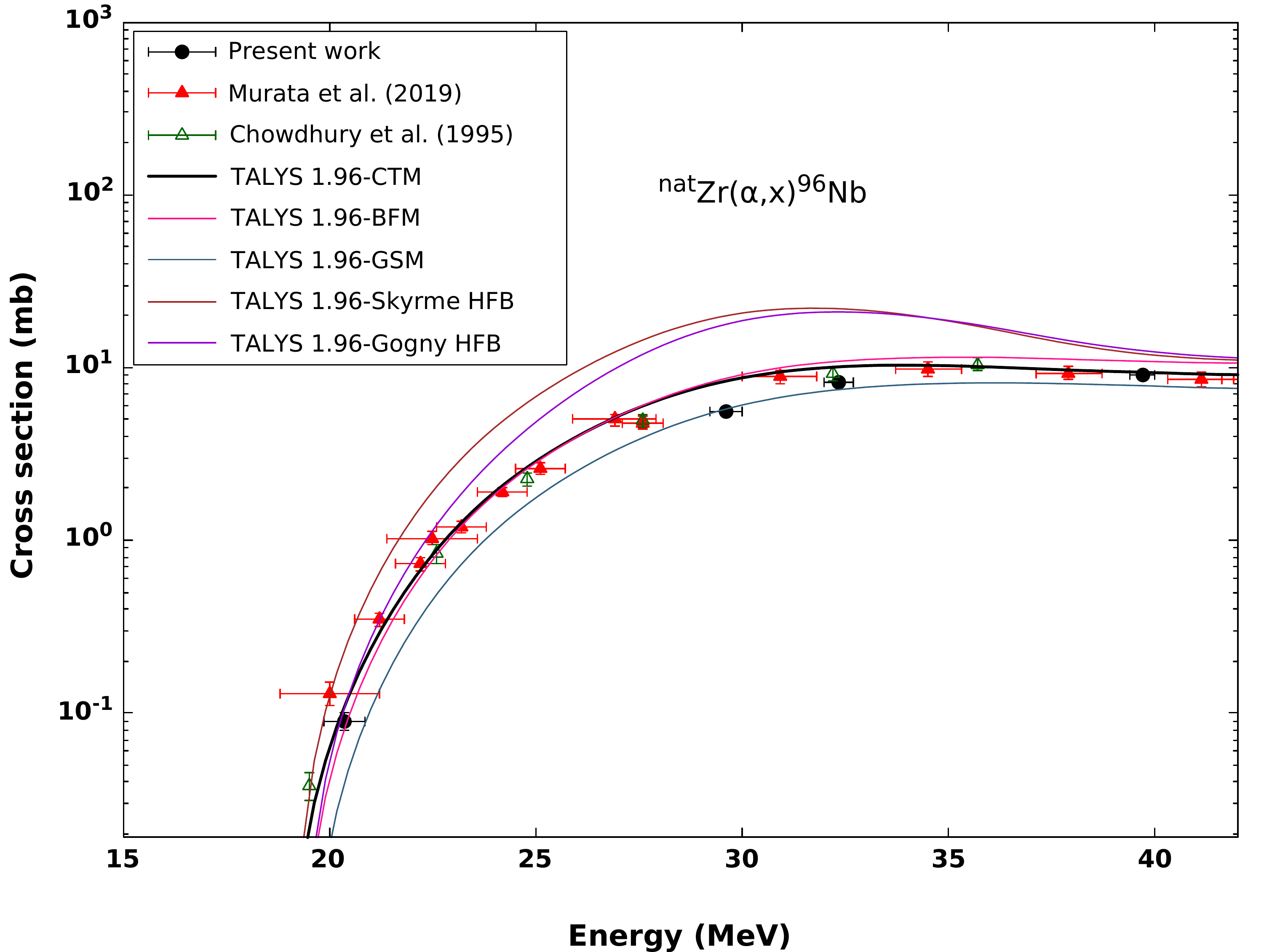}
\caption{Experimentally measured excitation functions of the reaction  {$^{nat}Zr(\alpha,x)^{96}Nb$}  with TALYS calculations.}\label{96Nb}
\end{figure}
 The major contributing channels for $^{nat}Zr(\alpha,x)^{96}Nb$ reaction for energies upto 40 MeV are {$^{94}Zr(\alpha,d)^{96}Nb$} and {$^{94}Zr(\alpha,pn)^{96}Nb$}. The gamma energies of 778.224 keV, 568.871 keV, 1091.349 keV, and 460.04 keV are considered for the determination of cross sections. Fig. \ref{96Nb}  shows the measured cross sections along with literature data \cite{chowdhury, MUrata} and theoretical calculations. Data by Murata et al. and Chowdhury et al. are in good agreement with present measurements and the Constant temperature model of level density gives the best fit with measured data.
\subsection{{$^{nat}Zr(\alpha,x)^{95m,g}Nb$} reaction}
\begin{figure}[h]
\centering
\includegraphics[scale=0.23]{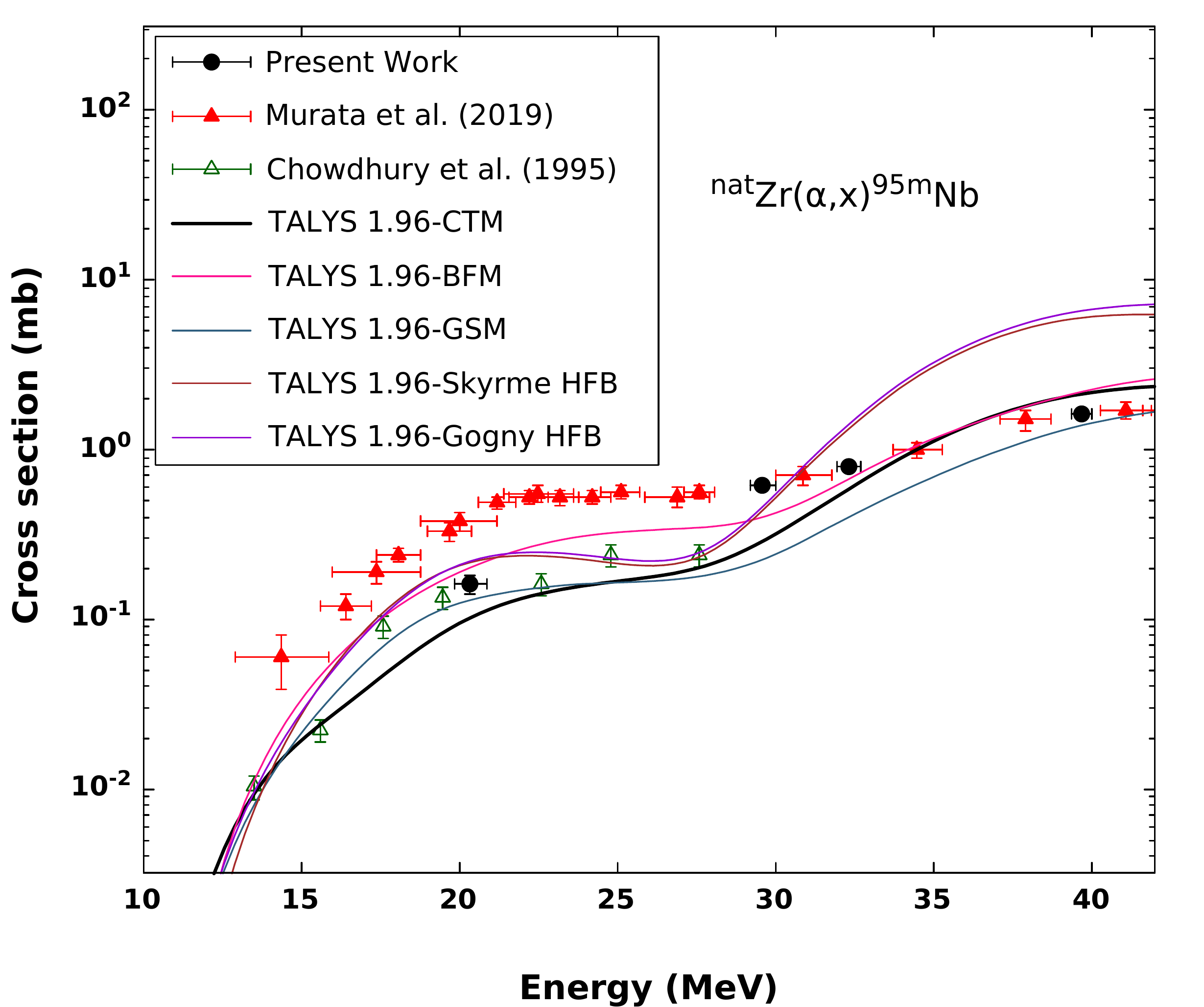}
\caption{Experimentally measured excitation functions of the reaction {$^{nat}Zr(\alpha,x)^{95m}Nb$} with TALYS  calculations.}\label{95Nb2}
\end{figure}
\begin{figure}[h]
\centering
\includegraphics[scale=0.23]{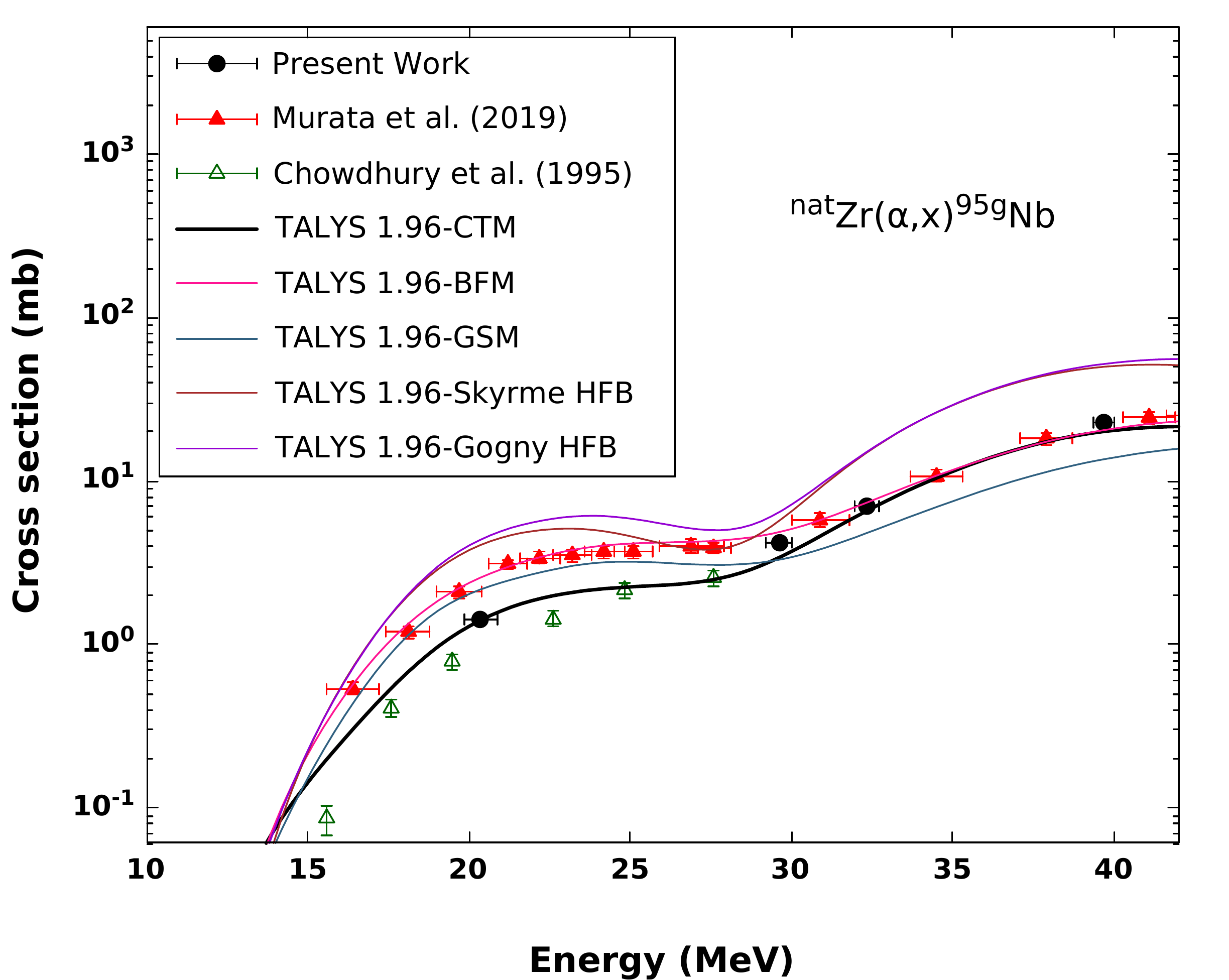}
\caption{Experimentally measured excitation functions of the reaction {$^{nat}Zr(\alpha,x)^{95g}Nb$} with TALYS calculations.}\label{95Nb1}
\end{figure}
	The radionuclide $^{95}Nb$ has an isomeric state $^{95m}Nb$ with a shorter half life of 3.6 days, and a ground state $^{95g}Nb$ with a half life of 34.991 days. The major contributing reaction channels are {$^{92}Zr(\alpha,p)$,} {$^{94}Zr(\alpha,t)$},  {$^{94}Zr(\alpha,nd)$} and {$^{94}Zr(\alpha,2np)$}. The production cross sections of the $^{nat}Zr(\alpha,x)^{95m}Nb$ and $^{nat}Zr(\alpha,x)^{95g}Nb$ reactions are measured for a photo peak energy 235.69 keV and 765.803 keV respectively.  The TALYS calculations with the Constant temperature model of level density shows good agreement with measurements and are shown in Fig. \ref{95Nb2} and Fig. \ref{95Nb1} respectively along with the previously measured data \cite{chowdhury,MUrata}.
\subsection{{$^{nat}Zr(\alpha,x)^{92m}Nb$} reaction}
\begin{figure}[h!]
\centering
\includegraphics[scale=0.23]{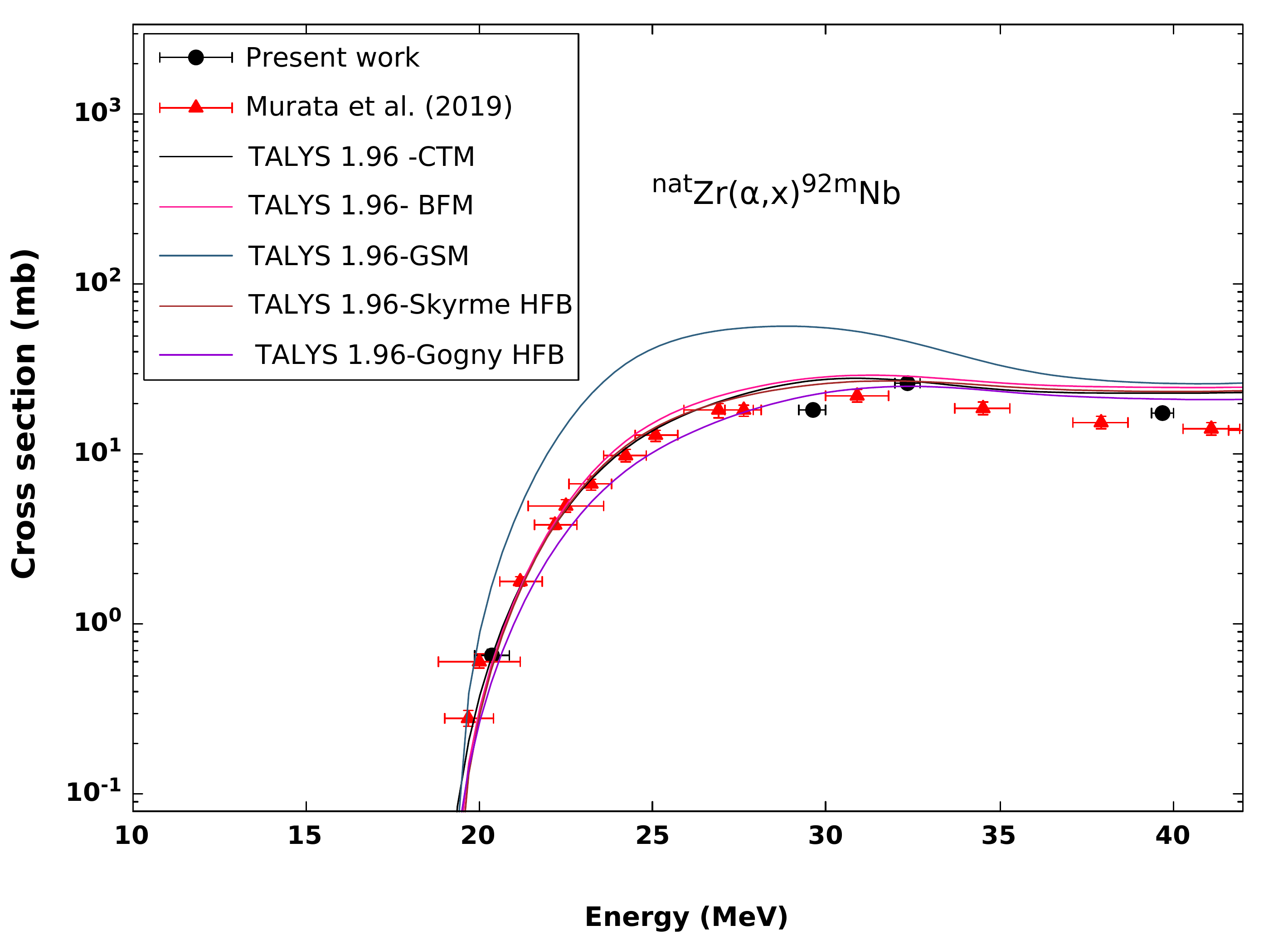}
\caption{Experimentally measured excitation functions of the reaction $^{nat}Zr(\alpha,x)^{92m}Nb$ with TALYS calculations.}\label{92mNb}
\end{figure} 

	The radionuclide $^{92}Nb$ has an isomeric state $^{92m}Nb$ with a shorter half life of 10.15 days, and a ground state $^{92g}Nb$ with a half life of $ 3.47 \times 10^7 $ years. Production cross sections for {{$^{nat}Zr(\alpha,x)^{92g}Nb$} reaction can not be measured since the half life of the residual nucleus is very long. Production of $^{92m}Nb$ radionuclide comes from the $^{90}Zr(\alpha,np)$ and $^{91}Zr(\alpha,2np)$ channels and measured cross sections for the gamma energy 934.44 keV are shown in Fig. \ref{92mNb}. TALYS calculations with level density models except for the GSM model of level density show a similar trend and agree well with present data.  
 \subsection{{$^{nat}Zr(\alpha,x)^{90}Nb$} reaction}
\begin{figure}[h!]
\centering
\includegraphics[scale=0.21]{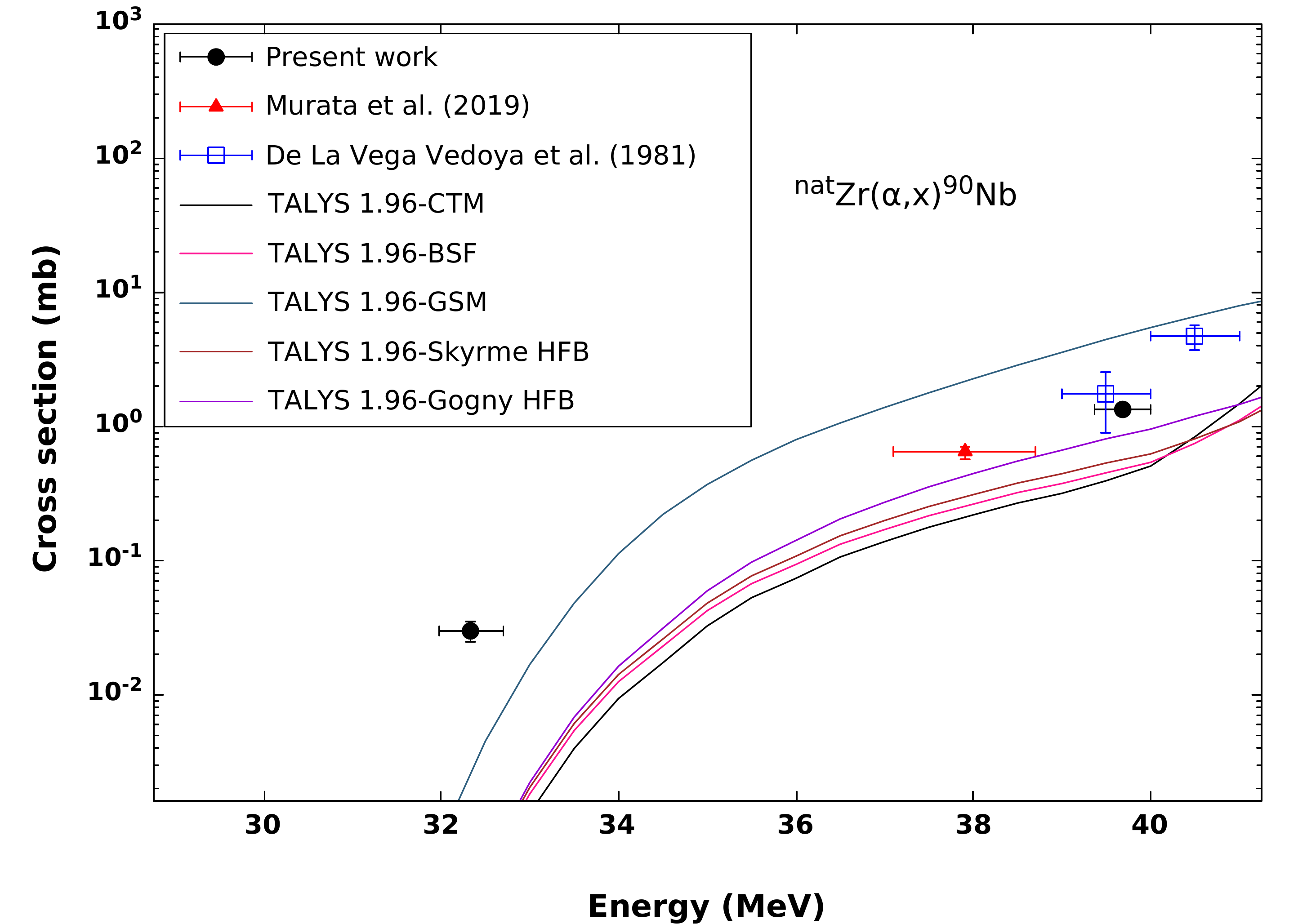} 
\caption{Experimentally measured excitation functions of the reaction $^{nat}Zr(\alpha,x)^{90}Nb$ with TALYS calculations.}\label{90Nb}
\end{figure}
{Production cross sections for $^{90}Nb$ in the $^{nat}Zr(\alpha,x)^{90}Nb$ reactions are measured for the gamma energy 1129.224 keV and are shown in Fig. \ref{90Nb}, and is being reported first time in the interested energy domain. Major contribution comes through the channel {$^{90}Zr(\alpha,tn)^{90}Nb$} in the interested energy domain. The threshold energy for other contributing channels is above the energy of  34.18 MeV. TALYS calculations with the GSM model give close agreement with the measurement at the lowest energy and all level density models except the GSM model show agreement with measurements at higher energies.

\subsection{{$^{nat}Zr(\alpha,x)^{89g}Zr$} reaction}
\begin{figure}[h]
\centering
\includegraphics[scale=0.25]{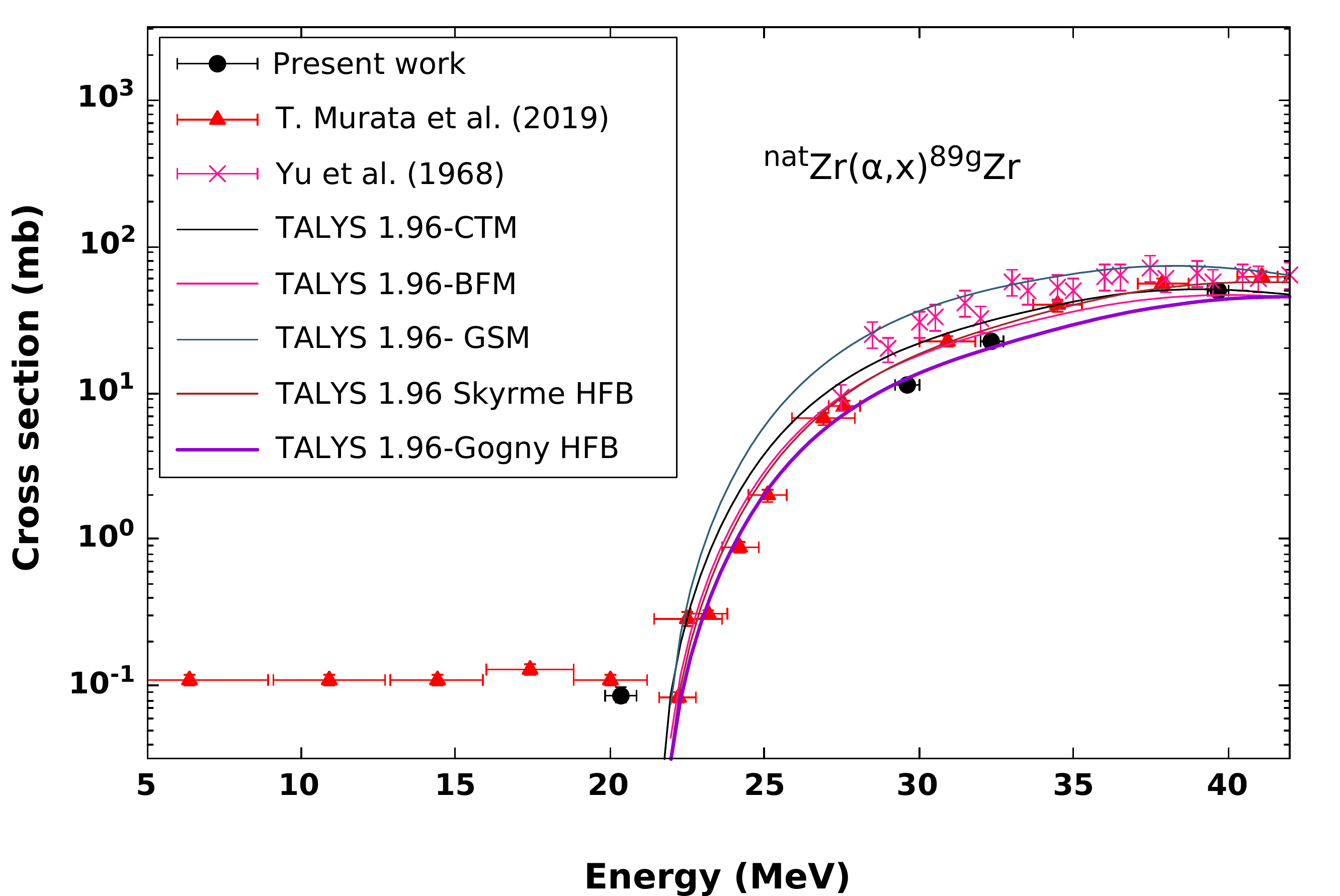} 
\caption{Experimentally measured excitation functions of the reaction $^{nat}Zr(\alpha,x)^{89g}Zr$ with TALYS calculations.}\label{89gZr}
\end{figure}
	The radionuclide $^{89}Zr$ has an isomeric state $^{89m}Zr$ with a shorter half life of 4.16 minutes, and a ground state $^{89g}Zr$ with a half life of 78.41 hours. Production cross sections for {{$^{nat}Zr(\alpha,x)^{89g}Zr$} reaction can not be measured since the half life of $^{89m}Zr$ nucleus is very short. The major contributing channels are {$^{90}Zr(\alpha,\alpha n)$}, {$^{91}Zr(\alpha,\alpha 2n)$}, and {$^{92}Zr(\alpha,\alpha 3n)$}. Production cross sections of radionuclide $^{89g}Zr$ through $^{nat}Zr(\alpha,x)^{89g}Zr$ reaction for gamma energy of 909.04 keV are shown in Fig. \ref{89gZr} along with measurements by Murata et al. and Yu et al. \cite{Yu}. TALYS calculations with the Gogny HFB model of level density shows good agreement with present measurements.	
\section{Determination of isomeric cross section ratios}
	Isomeric Cross section Ratio(ICR) is defined as the ratio of the cross section for the population of the isomeric state to the total cross section for the production of the particular isotope. ICR has been experimentally determined for $^{95}Nb$ and theoretically estimated for residual nuclei, where there is either the isomeric state or ground state or both are measurable. Theoretical estimation is based on the parameters that best fit the measured excitation functions. From the measured cross sections of $^{nat}Zr(\alpha,x)^{95g}Nb$ and $^{nat}Zr(\alpha,x)^{95m}Nb$ reactions, the isomeric cross section ratio for $^{nat}Zr(\alpha,x)^{95}Nb$ reaction has been determined. To the best of our knowledge, the ICR for this channel is experimentally measured and reported for the first time. Experimentally measured isomeric cross section ratios for the production of $^{95}Nb$  are listed in Table \ref{Table4} and plotted in Fig. \ref{icr} along with theoretical estimation using TALYS. 
\begin{table}[h!]
\small
\caption{The isomeric cross section ratios for $^{nat}Zr(\alpha,x)^{95g,m}Nb$ reaction.}\label{Table4}
\centering
\begin{tabular}{lr}
\hline
Energy(MeV) & {\centering Isomeric cross section ratio} \\
\hline
$20.4 \pm 0.5$ &  $ 0.1\pm 0.02 $\\
$29.6 \pm 0.4$ &  $ 0.13\pm 0.01 $\\
$32.3 \pm 0.4$ &  $ 0.1 \pm 0.01 $\\
$39.7 \pm 0.3$ &  $ 0.07 \pm 0.01$\\         
\hline
\end{tabular}
\end{table}
\begin{figure}[h!]
\centering
\includegraphics[scale=0.3]{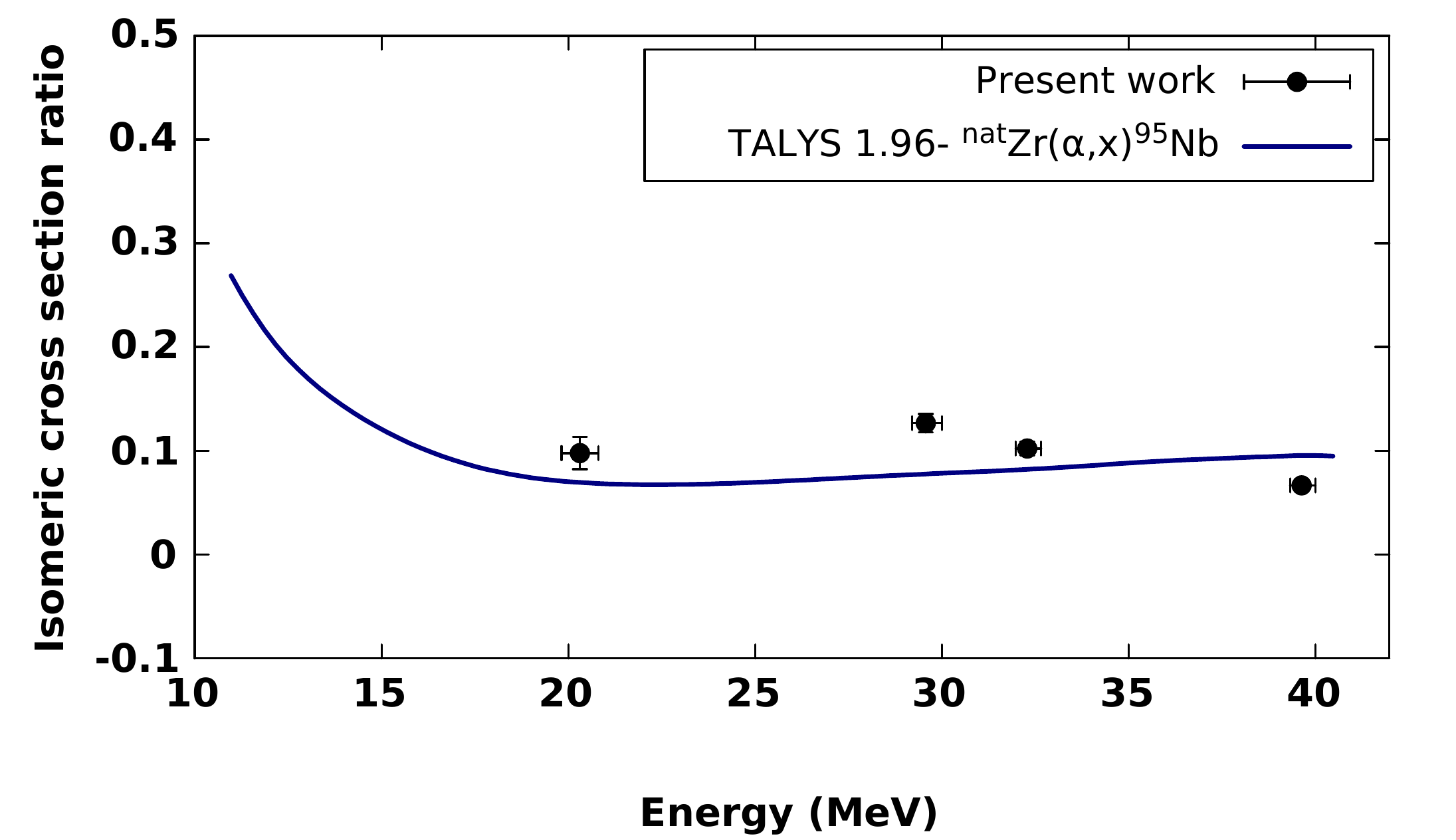}
\caption{Experimentally measured isomeric cross section ratios of the reaction $^{nat}Zr(\alpha,x)^{95g,m}Nb$ with TALYS calculations.}\label{icr}
\end{figure}
Isomeric cross section ratio shows a steady decreasing trend from 0.2 to 0.1 up to an energy of 20 MeV and then it is saturated to the value of 0.1, which implies that  isomeric state with spin and parity of ${1/2}^-$ at energy 235.69 keV having a relative population of 10 \% and the ground state with higher spin state(${9/2}^+$) is the more favourable state. 
 
	Similarly from the measured cross sections of the {$^{nat}Zr(\alpha,x)^{93m}Mo$}, {$^{nat}Zr(\alpha,x)^{92m}Nb$}, and {$^{nat}Zr(\alpha,x)^{89g}Zr$} reactions, ICR for $^{93}Mo$, $^{92}Nb$, and $^{89}Zr$ isotopes are determined using TALYS and are plotted in Fig. \ref{estimated}.
\begin{figure}[h]
\centering
\includegraphics[scale=0.27]{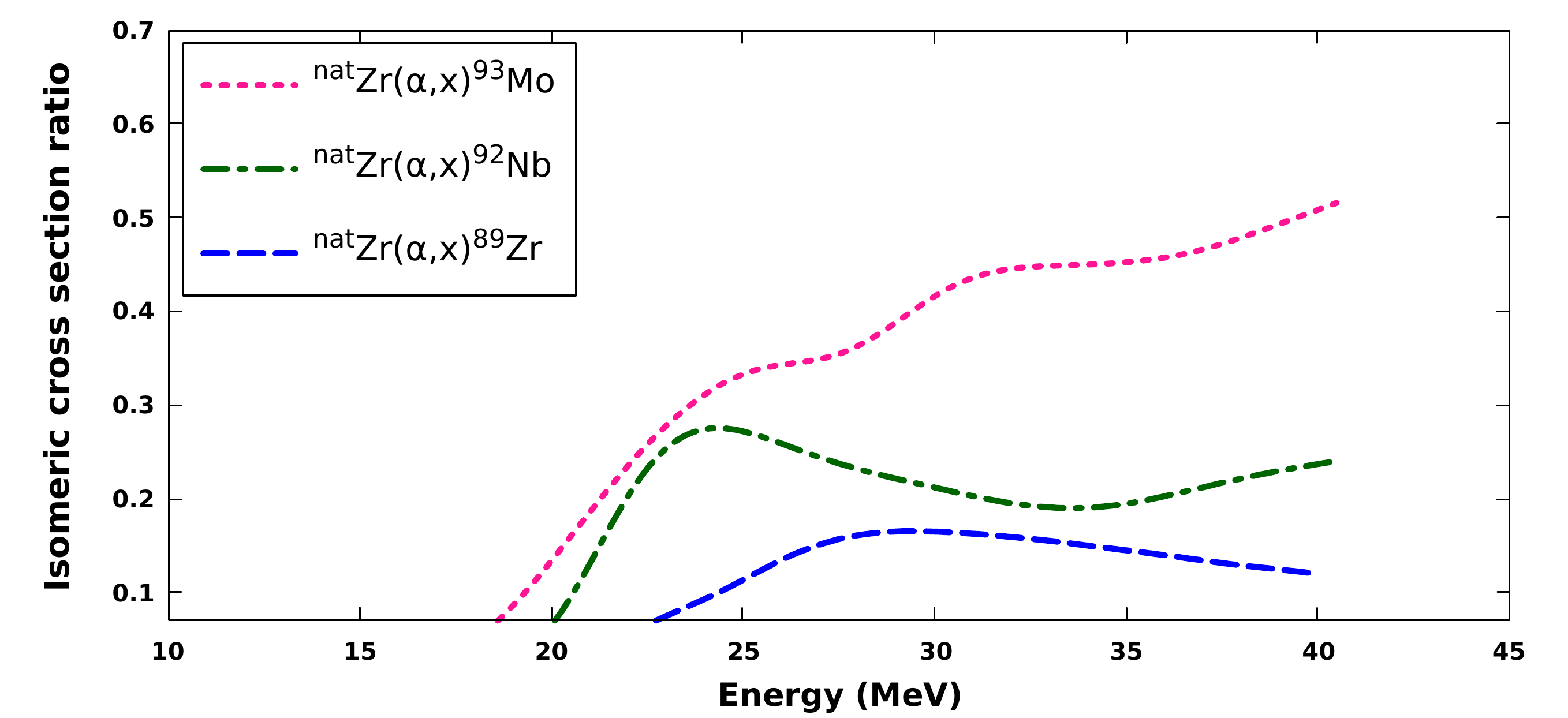} 
\caption{Theoretically estimated ICR for the  {$^{nat}Zr(\alpha,x)^{93}Mo$}, {$^{nat}Zr(\alpha,x)^{92}Nb$}, and {$^{nat}Zr(\alpha,x)^{89}Zr$} reactions.}\label{estimated}
\end{figure}	
ICR plot shows an increasing trend with alpha energy for  $^{93}Mo$ nuclei and a decreasing trend for  $^{92}Nb$ and $^{89}Zr$ nuclei after an initial increase. The increase in excitation energy results in the production of residual nuclei with higher angular momentum, hence the population at the higher spin state is generally favoured \cite{MUHAMMEDSHAN20189,MUHAMMEDSHAN2020}. However at lower excitation energies the ground state is populated first, irrespective of the spin state. This trend is also deserved in the case of  $^{89}Zr$, $^{121}Te$ and $^{197}Hg$ nuclei produced through proton induced reactions \cite{PhDThesis}.

\section{Estimation of neutron production cross sections}
 \begin{figure}[h!]
  \centering
  \includegraphics[scale=0.3]{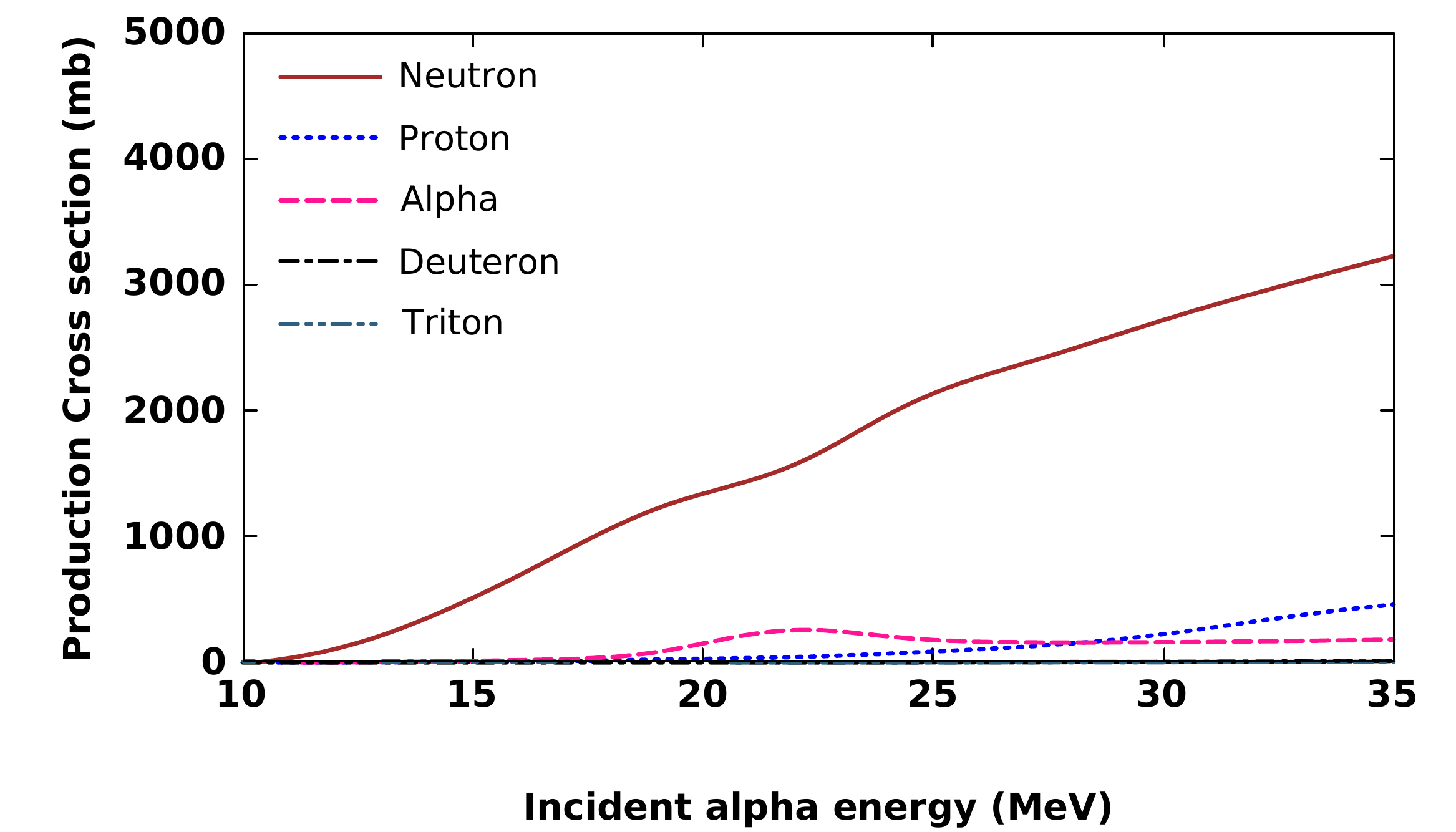}
  \caption{TALYS calculation of total production cross section for n, p, d, t and alpha particles in $^{nat}Zr(\alpha,x)$ reaction. }\label{particle_production}
  \end{figure} 
  
	The alpha induced reaction on natural Zirconium leads to the production of different residual nucleus through various possible channels, depending on the threshold value of that particular reaction channel. From the activation analysis, we can observe only the residual nuclei having half life in the order of cooling time and counting time, that is in a few minutes to a few days. However the reaction results in the production of the residual nucleus having half life shorter than and longer than the observed residues also. There will be a production of neutrons and other charged particles from all these channels. By considering the overall possible reaction channels, the production cross sections for neutron, proton, deuteron, triton, and alpha particles are estimated using the TALYS 1.96 code and are shown in Fig. \ref{particle_production}. The Constant temperature model of level density is considered for these calculations since it gives over all best fit with measured cross sections.

	From the results, it is found that the neutron production cross sections are in several barns and much higher compared to other particle production cross sections. It results in additional neutron contribution to the reactor environment and which affects the criticality. Further, the charged particles thus produced may cause the production of gas within the structural material causing rupture, and thereby damaging the structural material. Hence, interaction of alpha particles with structural material has to be properly considered for the safe operation of the reactor.
\section{Conclusion}
The present study highlights the significant effect of $\alpha$ particles produced in the fission process, producing various radio isotopes {$^{99}Mo$}, {$^{93m}Mo$}, {$^{96}Nb$}, { $^{95g}Nb$}, {$^{95m}Nb$}, {$^{92m}Nb$},  {$^{89g}Zr$}  and {$^{90}Nb$}  and  a much higher neutron production cross sections to the reactor domain thereby effecting the criticality of the reactor. Moreover, from the isomeric cross section ratio calculations in the energy range 15-40 MeV, provide important information on spin population of progressed reaction. ICR depends entirely on the excitation energy and relative spin state of the residual nucleus.

\section{Acknowledgment}
The authors extend their sincere thanks to  DAE-BRNS and  UGC-DAE CSR for providing financial support.
The one of the author Vafiya Thaslim T T thankful to CSIR, India for providing the financial support under the scheme of JRF. The authors acknowledge the support provided by Dr. Chandana Battacharya, Nuclear Physics Division BARC, VECC, A A Mallik ACD division, BARC, VECC and all the staff members  associated with Cyclotron facility.


\bibliography{sn-bibliography}


\end{document}